\documentclass[man,floatsintext]{apa7}

\usepackage[american]{babel}
\usepackage{csquotes}
\usepackage{microtype}
\usepackage{mathtools}
\usepackage{amssymb}
\usepackage{booktabs}
\usepackage{array}
\usepackage{threeparttable}
\usepackage{graphicx}
\usepackage{enumitem}
\usepackage{xcolor}
\usepackage{tikz}
\usetikzlibrary{positioning,arrows.meta,fit}
\usepackage{hyperref}
\usepackage[nameinlink,noabbrev]{cleveref}
\usepackage[style=apa,sortcites=true,sorting=nyt,backend=biber]{biblatex}

\newcommand{\Anx}{\ensuremath{L}}
\newcommand{\Avo}{\ensuremath{U}}
\newcommand{\Sec}{\ensuremath{S}}
\newcommand{\MentAff}{\ensuremath{M}}
\newcommand{\Flex}{\ensuremath{F}}

\hypersetup{
  pdftitle={Beyond Anxiety and Avoidance: A Dynamic Feedback-Regulation Framework for Adult Romantic Attachment},
  pdfauthor={Mateusz Pomianek},
  colorlinks=true,
  linkcolor=black,
  citecolor=black,
  urlcolor=blue
}

\title{Beyond Anxiety and Avoidance: A Dynamic Feedback-Regulation Framework for Adult Romantic Attachment}
\shorttitle{Dynamic Adult Attachment Regulation}

\authorsnames{Mateusz Pomianek}
\authorsaffiliations{Rzeszów University of Technology}

\abstract{Adult attachment is commonly represented by attachment anxiety and avoidance, but these person-level dimensions do not by themselves specify how momentary relational conditions are perceived, compared with desired conditions, translated into action, and updated after interpersonal feedback. We present the Attachment Regulation and Control (ARC) framework as an integrative feedback-regulation account that makes these temporal assumptions explicit and empirically testable. In ARC, control refers to feedback-based regulation toward relational goals in the control-theoretic sense, not interpersonal dominance, coercion, or a generalized need for control. ARC does not claim a novel control-system architecture. Instead, it synthesizes established control-theoretic attachment ideas, secure-base representations, perceived partner responsiveness, relational discrepancy, dyadic regulation, and recovery into a parsimonious sequence for adult romantic relationships: perceived relational state, desired relational state, discrepancy, regulatory action, partner feedback, state updating, and recovery. The framework treats the discrepancy score itself as a falsifiable modeling restriction by comparing it with perceived-state-only, unconstrained perceived-plus-desired-state, and asymmetric discrepancy alternatives. Attachment anxiety and avoidance form core regulation moderators. Secure-base representation is a candidate requiring incremental prediction of recovery outcomes. Mentalized affectivity and regulation flexibility remain optional moderators, while adult disorganization is an independent extension outcome.

}


\begin{document}
\maketitle
\section{Introduction}

Adult attachment research is strongly organized around attachment anxiety and attachment avoidance \parencite{fraley2000irt,mikulincer2016attachment,li2012meta}. These dimensions describe stable individual differences in hyperactivating and deactivating tendencies, but they do not uniquely specify the micro-temporal process through which a person interprets a relational event, compares the current condition with a desired condition, acts, receives a partner response, and recovers or remains dysregulated.

The need for a dynamic account is not itself new. Bowlby's theory was explicitly goal directed and feedback regulated \parencite{bowlby1982attachment}; control-theory analyses were applied to adolescent attachment and emotion regulation decades ago \parencite{kobak1993control}; computational implementations have formalized attachment set-goals and feedback \parencite{petters2019control,gagliardi2022control,stevens2009dynamic}; and adult relationship theories already describe risk regulation, partner responsiveness, coregulation, and partner buffering of insecurity \parencite{murray2006risk,sbarra2008coregulation,simpson2014buffering,overall2022buffering}. We refer to the proposed framework as Attachment Regulation and Control (ARC). Here, control denotes feedback-based regulation toward relational goals in the control-theoretic sense; it does not denote interpersonal dominance, coercion, or a generalized need for control. ARC therefore does not present feedback control as a new idea.

The problem addressed here is narrower and empirical. Existing ideas are distributed across different levels of analysis and are rarely joined in a measurement architecture that can be tested as a temporally ordered process in adult romantic relationships. ARC asks whether person-level attachment orientations predict how repeatedly measured relational discrepancies are translated into action, how partner feedback updates subsequent relational state, and whether a positive secure-base representation adds recovery information beyond established insecurity dimensions.

This framing also limits claims about security and disorganization. Secure-base script research provides positive representational content for security beyond merely low anxiety and avoidance \parencite{waters2006workingmodels,waters2019securebase,waters2025securebase}. This justifies prespecifying positive secure-base representation as a candidate construct, but does not yet justify making it a required ARC component because its incremental value for recovery beyond anxiety and avoidance remains unestablished. Adult disorganization, by contrast, remains heterogeneous in definition and measurement \parencite{pollard2023disorganized}. ARC therefore treats recovery as a dynamic outcome to be measured, positive security as a candidate to be tested against it, and disorganization as an independent validation criterion rather than a quantity generated by the model.

ARC does not require a particular computational controller family. P, PI, PID, state-space, active-inference, reinforcement-learning, and related formulations can be treated as competing mechanistic or observation models when their identifying predictions are available. The psychological framework is instead defined by measurable relations among perceived state, desired state, discrepancy, action, feedback, state updating, and recovery. This separation allows the framework to remain testable even when several latent mechanisms are observationally equivalent.

\subsection{Contribution and scope}

The strongest defensible contribution of ARC is integration and empirical specification, not invention of its individual components. The article contributes: (a) a parsimonious temporal sequence for adult romantic attachment regulation; (b) an explicit distinction among required core variables, prespecified candidates that must earn promotion, and optional extensions; (c) falsifiable predictions that can be tested with intensive longitudinal and dyadic data; and (d) criteria for simplifying, withholding, or rejecting unsupported components.

The article is theoretical and methodological. It contains no human validation data. Its central claim remains provisional until the proposed temporal variables show incremental and out-of-sample value in human observations.

\section{Theoretical Foundations}

\subsection{Control-system attachment is prior theory, not ARC novelty}

Bowlby's attachment system was formulated as goal directed and feedback regulated \parencite{bowlby1982attachment}. Kobak and colleagues later used control theory to analyze attachment strategies and emotion regulation during mother--teen problem solving \parencite{kobak1993control}. Petters reviewed the control-system lineage and its computational implications \parencite{petters2019control}. ARC adopts this lineage rather than claiming to originate it.

\subsection{Computational and dynamical precedents}

Stevens and Zhang modeled infant attachment as a dynamic system with explicit state evolution and recovery-like behavior \parencite{stevens2009dynamic}. Gagliardi implemented attachment as a multidimensional control system with representational set-goals and agent interactions \parencite{gagliardi2022control}. Olar and colleagues used an adaptive network model for anxious--avoidant relationship dynamics \parencite{olar2026relationship}. These precedents show that targets, state updating, feedback, and dyadic dynamics are already computationally plausible.

Their existence changes the novelty question. ARC should not be evaluated by whether it can reproduce another simulated attachment trajectory. It should be evaluated by whether its proposed variables can be measured repeatedly in people and whether the resulting temporal model adds information beyond static attachment scores and simpler time-series baselines.

\subsection{Risk regulation, responsiveness, and coregulation}

Adult relationship regulation also has established psychological precedents. Risk-regulation theory describes how dependence and perceived acceptance shape decisions to seek connection or protect the self \parencite{murray2006risk}. Coregulation and partner-buffering accounts emphasize that adult attachment, distress, separation, and recovery unfold through coupled interpersonal regulation rather than isolated intrapersonal control \parencite{sbarra2008coregulation,simpson2014buffering,overall2022buffering}. ARC therefore treats partner feedback as a measured component of the process rather than background noise.

\subsection{Closeness and autonomy are not assumed to be opposites}

ARC distinguishes experienced autonomy from interpersonal distance. Close-relationship research shows that autonomy and relatedness can be jointly satisfied rather than occupying opposite ends of one continuum \parencite{kluwer2020autonomy}. This supports a possible multidimensional relational-state representation while stopping short of establishing which coordinates ARC requires. The scalar closeness pilot therefore remains the first test; a closeness--autonomy extension must earn its added complexity empirically.

\subsection{Positive security as a candidate core construct}

Secure-base representations encode expectations that support can be available and effective and that exploration or constructive activity can resume \parencite{waters2006workingmodels,waters2019securebase,waters2025securebase}. This provides a strong theoretical reason to test positive secure-base representation, \Sec, as distinct from simply reversing \Anx\ and \Avo. It does not, however, establish that \Sec\ is necessary for the ARC architecture or that it predicts recovery beyond dimensional insecurity. ARC therefore prespecifies \Sec\ as a candidate core construct rather than a required component. Recovery is treated as an observable dynamic outcome against which this candidate can earn or fail to earn incremental validity.

\subsection{Candidate moderators rather than additional core constructs}

Mentalized affectivity concerns identifying, processing, and expressing affective states \parencite{greenberg2017mentalized}. Emotion-regulation flexibility concerns context-sensitive adaptation \parencite{mosannenzadeh2024flexibility}. Both are relevant to ARC, but their role is deliberately limited. Mentalized affectivity is a candidate moderator of relational-state inference quality; flexibility is a candidate moderator of context- and feedback-sensitive policy adaptation. Neither is required for the core feedback loop to be defined.

\subsection{The empirical gap}

The remaining gap is a measurement problem: can the sequence from perceived and desired relational conditions to discrepancy, action, partner feedback, and recovery be recovered in adult romantic relationships using intensive longitudinal data? ARC is organized around this question.

\section{The Attachment Regulation and Control Framework}

\subsection{A parsimonious core}

ARC distinguishes person-level modulators from momentary process variables.

\textbf{Person-level core.} Attachment anxiety (\Anx) and avoidance (\Avo) are the current required person-level dimensions and candidate moderators of relational regulation.

\textbf{Prespecified candidate core.} Positive secure-base representation (\Sec) is not required to instantiate ARC. It is retained as a theoretically motivated candidate whose proposed dynamic correlate is effective recovery. H3 functions as its promotion test: \Sec\ should enter the person-level core only if it demonstrates reproducible incremental validity for recovery beyond \Anx\ and \Avo.

\begin{table}[htbp]
\caption{Construct Status in ARC 0.9.0}
\label{tab:construct-status}
\centering
\small
\setlength{\tabcolsep}{3pt}
\renewcommand{\arraystretch}{0.95}
\begin{threeparttable}
\begin{tabular}{p{0.15\linewidth}p{0.17\linewidth}p{0.29\linewidth}p{0.27\linewidth}}
\toprule
Construct & Current status & Role & Retention or promotion rule \\
\midrule
\(\Anx,\Avo\) & Core person-level reference dimensions & Established attachment dimensions; H2 tests whether they moderate channel-specific regulation & Retain as attachment descriptors; remove ARC-specific dynamic moderation if H2 does not generalize \\
\(\Sec\) & Prespecified candidate core & Positive secure-base representation tested against recovery & Promote only after incremental value beyond \(\Anx,\Avo\) on a preregistered primary recovery outcome, out-of-sample improvement, and replication \\
\(\MentAff,\Flex\) & Optional extensions & Candidate pathway-specific moderators & Retain only with discriminant and incremental validity for their assigned pathways \\
\(Z^{\mathrm{obs}}\) & External validation criterion & Independently measured adult disorganization & Never define from ARC dynamics alone; use to test incremental validity of H5 indicators \\
\(C\) & Separate psychometric program & Candidate relational-control propensity & Not part of ARC unless independently validated and theoretically justified in a later model \\
\bottomrule
\end{tabular}
\begin{tablenotes}[flushleft]
\footnotesize
\item Note. Status refers to the ARC model architecture, not to whether the underlying construct exists in the broader attachment literature.
\end{tablenotes}
\end{threeparttable}
\end{table}

\textbf{Momentary core.} At observation \(t\), the person reports or otherwise yields an estimate \(\hat{\mathbf r}_{i,t}\) of the current relational state and a desired state \(\mathbf r^\ast_{i,t}\) on matched coordinates. Their discrepancy is
\begin{equation}
\mathbf e_{i,t}=\mathbf r^\ast_{i,t}-\hat{\mathbf r}_{i,t}.
\label{eq:core-discrepancy}
\end{equation}
The subsequent regulatory action is \(\mathbf a_{i,t+1}\), followed by measured partner or environmental feedback \(\mathbf p_{i,t+1}\), an updated relational state, and recovery or persistence.

The core sequence is therefore:
\[
\hat{\mathbf r}_{t},\mathbf r^\ast_t
\rightarrow
\mathbf e_t
\rightarrow
\mathbf a_{t+1}
\rightarrow
\mathbf p_{t+1}
\rightarrow
\hat{\mathbf r}_{t+1}
\rightarrow
\mathrm{recovery}.
\]

\subsection{Relational state and target}

Relational state can be scalar in an initial pilot or multidimensional when the data justify it. Candidate coordinates include perceived responsiveness, availability, closeness, or autonomy. The desired state must be measured on the same coordinates. Autonomy here means experienced volition rather than low closeness or overt distancing. Evidence from close-relationship research indicates that autonomy and relatedness can be jointly satisfied, so a person may want both closeness and autonomy rather than trading one mechanically against the other \parencite{kluwer2020autonomy}.

A natural two-coordinate extension is
\[
\mathbf r_{i,t}=
\begin{bmatrix}
r^{\mathrm{close}}_{i,t}\\
r^{\mathrm{aut}}_{i,t}
\end{bmatrix},
\qquad
\mathbf e_{i,t}=
\begin{bmatrix}
r^{\ast,\mathrm{close}}_{i,t}-\hat r^{\mathrm{close}}_{i,t}\\
r^{\ast,\mathrm{aut}}_{i,t}-\hat r^{\mathrm{aut}}_{i,t}
\end{bmatrix}.
\]
This representation permits simultaneously high desired closeness and high desired autonomy. It does not assume that the two coordinates are opposite ends of one axis. Whether they behave as approximately separable regulatory loops or require cross-coordinate coupling is an empirical question represented by the off-diagonal elements of \(\mathbf B_{e,i}\) in Equation~\ref{eq:minimal-action}. A two-coordinate extension therefore requires separate matched state and target ratings and its own measurement and design evaluation. It is not an additional confirmatory variable in the current scalar H1 pilot. ARC does not assume one universal set point; targets may differ between people, partners, contexts, and episodes.

\subsection{Minimal action model}

The first empirical model should be deliberately modest:
\begin{equation}
\mathbf a_{i,t+1}
=
\boldsymbol{\alpha}_i
+
\mathbf B_{e,i}\mathbf e_{i,t}
+
\mathbf \Phi_i\mathbf a_{i,t}
+
\mathbf B_{c,i}\mathbf c_{i,t}
+
\boldsymbol{\epsilon}_{i,t+1}.
\label{eq:minimal-action}
\end{equation}

Crucially, the discrepancy term is a testable restriction rather than a mandatory preprocessing step. For a scalar state and target, the unconstrained target-aware model can be written as
\begin{equation}
a_{i,t+1}
=
\alpha_i
+
\beta_{\hat r}\hat r^{\mathrm{WP}}_{i,t}
+
\beta_{r^\ast}r^{\ast,\mathrm{WP}}_{i,t}
+
\phi_i a_{i,t}
+
\boldsymbol{\beta}^{\top}_c\mathbf c_{i,t}
+
\epsilon_{i,t+1}.
\label{eq:unconstrained-target-state}
\end{equation}
Substituting \(e^{\mathrm{WP}}_{i,t}=r^{\ast,\mathrm{WP}}_{i,t}-\hat r^{\mathrm{WP}}_{i,t}\) into Equation~\ref{eq:minimal-action} shows that the signed-discrepancy model imposes
\begin{equation}
\beta_{r^\ast}=\beta_e,
\qquad
\beta_{\hat r}=-\beta_e.
\label{eq:discrepancy-restriction}
\end{equation}
Thus M3 asserts more than predictive usefulness of state and target: it asserts that their within-person contributions are equal in magnitude and opposite in sign on the action scale. M2 directly relaxes this restriction, whereas M4 relaxes linear symmetry by estimating closeness-deficit and closeness-surfeit slopes separately. This nesting logic makes the difference-score representation empirically defeasible.

The first empirical hinge is therefore tested with direct, prespecified coefficient-level claims rather than a sequential predictive gate. H1-P1 asks whether desired relational state contributes to subsequent regulation beyond perceived state, context, and behavioral persistence in the unconstrained M2 model. H1-P2 asks whether the signed discrepancy predicts subsequent regulation in the preregistered direction in M3. The equality restriction in Equation~\ref{eq:discrepancy-restriction} is then evaluated separately through the contrast \(d_{\mathrm{sym}}=\beta_{r^\ast}+\beta_{\hat r}\), while M4 describes positive--negative discrepancy asymmetry with an estimate and confidence interval; the current H1 pilot has no confirmatory asymmetry threshold. Held-out prediction remains an important secondary validation layer, but failure of one restricted predictive approximation does not prevent evaluation of another prespecified representation. More complex history, nonlinear, state-space, switching, active-inference, reinforcement-learning, or P/PI/PID models are competitors, not assumptions.

\subsection{Attachment-dependent policy is channel specific}

For H2, regulatory action is represented as a vector rather than only by a signed summary:
\[
\mathbf a_{i,t}
=
\begin{bmatrix}
a^{\mathrm{approach}}_{i,t} \\
a^{\mathrm{distance}}_{i,t} \\
a^{\mathrm{internal}}_{i,t}
\end{bmatrix}.
\]
A channel-specific extension of Equation~\ref{eq:minimal-action} is
\begin{equation}
a^{(k)}_{i,t+1}
=
\alpha^{(k)}_i
+
\beta^{(k)}_e e^{\mathrm{WP}}_{i,t}
+
\gamma^{(k)}_L \left(\Anx_i e^{\mathrm{WP}}_{i,t}\right)
+
\gamma^{(k)}_U \left(\Avo_i e^{\mathrm{WP}}_{i,t}\right)
+
\phi^{(k)}_i a^{(k)}_{i,t}
+
\boldsymbol{\beta}^{(k)\top}_c\mathbf c_{i,t}
+
\epsilon^{(k)}_{i,t+1},
\label{eq:h2-policy}
\end{equation}
for \(k\in\{\mathrm{approach},\mathrm{distance},\mathrm{internal}\}\).

This formulation allows attachment dimensions to alter both response magnitude and response channel. It avoids assuming that higher avoidance is simply the negative end of the same behavioral axis as higher anxiety. The signed action score remains useful for H1 parsimony, but H2 is evaluated on the component channels as well.

The strongest preregisterable directional expectation concerns closeness deficit (\(e_t>0\)): higher \Anx\ should be associated with stronger overt proximity-seeking or reassurance-oriented responses, whereas higher \Avo\ should be associated with weaker overt approach and greater use of distancing or internal disengagement. Because attachment-related responses are conditional on relational context and attachment-system activation \parencite{campbell2011interactionist}, channel-specific moderation is more defensible than a universal ``high gain'' versus ``low gain'' interpretation.

\subsection{Feedback and state updating}

A minimal transition model is
\begin{equation}
\hat{\mathbf r}_{i,t+1}
=
f_i(
\hat{\mathbf r}_{i,t},
\mathbf a_{i,t+1},
\mathbf p_{i,t+1},
\mathbf c_{i,t+1}
)
+
\boldsymbol{\omega}_{i,t+1}.
\label{eq:state-update}
\end{equation}

Here \(\mathbf p_{i,t+1}\) denotes partner or environmental feedback available to the actor. In a single-participant design this may be the actor's perceived partner response. In a stronger dyadic design, perceived feedback should be separated from the partner's self-reported or independently coded behavior. The latter is not treated as an error-free objective measure, but as a second source that can distinguish actor perception from partner-reported or observed behavior. This distinction matters because perceived responsiveness is itself psychologically consequential and attachment related \parencite{ruan2020responsiveness,segal2016responsiveness}.

For partners \(A\) and \(B\), a dyadic extension can write the feedback received by \(A\) as
\[
\mathbf p_{A,t+1}
=
g_A\!\left(
\mathbf a_{B,t+1},
\hat{\mathbf r}_{B,t},
\mathbf c_{AB,t+1}
\right)
+
\boldsymbol{\nu}_{A,t+1},
\]
with an analogous equation for \(B\). Thus partner feedback is generated partly by the partner's regulatory action but need not equal that action one-to-one. A later APIM-style or dynamic dyadic model can estimate actor, partner, and reciprocal cross-lagged paths while keeping perceived responsiveness distinct from partner-reported behavior.

\subsection{Discrepancy reduction can occur through two routes}

Because
\begin{equation}
\mathbf e_{i,t+1}-\mathbf e_{i,t}
=
\left(\mathbf r^\ast_{i,t+1}-\mathbf r^\ast_{i,t}\right)
-
\left(\hat{\mathbf r}_{i,t+1}-\hat{\mathbf r}_{i,t}\right),
\label{eq:discrepancy-decomposition}
\end{equation}
a reduction in discrepancy can arise because perceived relational conditions move toward the target, because the target itself changes, or because both processes occur. ARC therefore does not equate discrepancy reduction with one psychological mechanism.

This decomposition is especially important for adult relationships. A person may restore closeness by eliciting partner responsiveness, or may adaptively revise a closeness goal after new information. Conversely, apparent recovery could reflect defensive target retreat rather than improved relational conditions. Empirical analyses should retain \(\hat{\mathbf r}_t\) and \(\mathbf r^\ast_t\) separately even when their difference is the primary predictor.

\subsection{Recovery as a multidimensional outcome}

Recovery is not defined by one exponential return coefficient. Following an attachment-relevant perturbation at \(t_0\), candidate preregistered quantities include
\begin{align}
B_i &= \sum_{t=t_0}^{t_1} |\mathbf e_{i,t}|\,\Delta t_{i,t},
\\
T^{\mathrm{rec}}_i
&=
\inf\left\{
t>t_0:
|\mathbf e_{i,t}|<\delta
\text{ for }m\text{ consecutive observations}
\right\},
\end{align}
together with overshoot, return to constructive activity, and the relative contribution of state change versus target change. \(B_i\) is discrepancy burden and \(T^{\mathrm{rec}}_i\) is time to a preregistered functional tolerance. These are parallel outcomes, not ingredients of a single unvalidated recovery index.

The H3 promotion claim is correspondingly specific: positive secure-base representation should predict more effective recovery beyond low \Anx\ and \Avo\ across preregistered recovery outcomes. Evidence should be reproducible and not depend on one selectively favorable metric. Until that criterion is met, \Sec\ remains outside the required person-level core.

\subsection{Positive security must earn promotion to the core}

ARC distinguishes \Sec\ from the absence of insecurity, but treats this distinction as a candidate empirical addition rather than a protected core assumption. The minimum H3 promotion comparison is therefore hierarchical:
\[
R_0
\subset
R_1=R_0+\Anx+\Avo
\subset
R_2=R_1+\Sec.
\]
\(R_0\) contains preregistered episode-level and design covariates. Promotion of \Sec\ to the ARC person-level core requires \(R_2\) to improve prediction of a preregistered primary recovery outcome relative to \(R_1\), with the increment surviving held-out or otherwise genuinely out-of-sample evaluation. A correlation between \Sec\ and recovery is insufficient if the same information is already contained in low \Anx\ and \Avo. Promotion should occur only after the incremental effect is independently replicated or supported by a second preregistered design; secondary recovery metrics provide convergence but cannot substitute for failure on the primary outcome.

A stronger extension adds proximal partner support or observed dyadic coping to form \(R_3\). This distinguishes a positive person-level secure-base representation from the support actually available in a given episode. If \Sec\ loses all incremental value once proximal support is included, it should not be promoted as an independent core construct; the theory should instead emphasize episode-level support or coregulation.

\subsection{Candidate moderators}

Mentalized affectivity (\MentAff) and emotion-regulation flexibility (\Flex) are placed outside the core. \MentAff\ is tested only as a candidate predictor or moderator of externally scored relational-state inference accuracy. \Flex\ is tested only as a candidate predictor or moderator of context-sensitive policy revision and effective feedback use. If they do not provide discriminant and incremental information, ARC does not require them.

\subsection{Disorganization as an external extension}

ARC does not define adult disorganization as high \Anx\ plus high \Avo, raw variability, or a simulator index. Candidate dynamic indicators such as poor recovery, context-residual inconsistency, persistent approach--avoidance conflict, and maladaptive switching must be validated against an independent disorganization criterion.

\begin{figure}[htbp]
\centering
\resizebox{\linewidth}{!}{\begin{tikzpicture}[
  node distance=11mm and 11mm,
  box/.style={draw, rounded corners, align=center, inner sep=5pt, minimum height=10mm},
  mod/.style={draw, dashed, rounded corners, align=center, inner sep=5pt},
  arrow/.style={-{Latex[length=2.1mm]}, thick},
  dashedarrow/.style={-{Latex[length=2.1mm]}, thick, dashed}
]
\node[box] (traits) {Core person level\\\(\Anx,\Avo\)};
\node[mod, below=of traits] (secure) {Candidate core\\\(\Sec\) via H3};
\node[box, right=of traits] (state) {Perceived + desired state\\\(\hat{\mathbf r}_t,\mathbf r^\ast_t\)};
\node[box, right=of state] (error) {Discrepancy\\\(\mathbf e_t\)};
\node[box, right=of error] (action) {Regulatory action\\\(\mathbf a_{t+1}\)};
\node[box, below=of action] (feedback) {Partner feedback\\\(\mathbf p_{t+1}\)};
\node[box, left=of feedback] (update) {Updated state\\\(\hat{\mathbf r}_{t+1}\)};
\node[box, left=of update] (recovery) {Recovery /\\persistence};
\node[mod, above=of error] (mods) {Optional moderators\\\(\MentAff,\Flex\)};

\draw[arrow] (state) -- (error);
\draw[arrow] (error) -- (action);
\draw[arrow] (action) -- (feedback);
\draw[arrow] (feedback) -- (update);
\draw[arrow] (update) -- (recovery);
\draw[arrow] (update.north) to[bend right=24] (state.south);
\draw[dashedarrow] (traits) -- (state);
\draw[dashedarrow] (traits.south) to[bend right=28] (action.west);
\draw[dashedarrow] (secure.east) to[bend right=18] (recovery.west);
\draw[dashedarrow] (mods) -- (state);
\draw[dashedarrow] (mods) -- (action);
\end{tikzpicture}}
\caption{Parsimonious ARC architecture. Anxiety and avoidance form the current person-level core and may modulate relational policy. Positive secure-base representation is a prespecified candidate for promotion to the core if it adds recovery information beyond anxiety and avoidance. Mentalized affectivity and emotion-regulation flexibility remain optional moderators.}
\label{fig:arc2-framework}
\end{figure}

\section{Empirical Evidence and Falsifiable Predictions}

\subsection{What is already established}

ARC builds on components that should not be presented as novel. Attachment anxiety and avoidance are established dimensional descriptions of adult attachment, and meta-analytic evidence links both dimensions to romantic relationship quality \parencite{fraley2000irt,mikulincer2016attachment,li2012meta}. Positive secure-base representations have an independent theoretical and measurement tradition \parencite{waters2006workingmodels,waters2019securebase,waters2025securebase}. Control-system interpretations of attachment predate ARC \parencite{bowlby1982attachment,kobak1993control,petters2019control}, and computational models already implement targets, state updating, feedback, and recovery-like dynamics \parencite{stevens2009dynamic,gagliardi2022control,olar2026relationship}. Adult relationship theories also provide established accounts of risk regulation, coregulation, and partner buffering of attachment insecurity \parencite{murray2006risk,sbarra2008coregulation,simpson2014buffering,overall2022buffering}.

The empirical contribution of ARC therefore depends on joining these strands in a single temporal measurement architecture and testing whether the proposed dynamic variables add information beyond simpler alternatives.

\subsection{Relational discrepancy has empirical precedent, but H1 remains under-tested}

The distinction between experienced and desired closeness has direct empirical precedent. Longitudinal work using actual and ideal closeness showed that smaller discrepancies were associated with better relational well-being and mental health, and that moving closer to one's ideal over time accompanied improved outcomes \parencite{frost2013closeness}. Dyadic response-surface analyses likewise showed that closeness discrepancy is consequential for relationship quality, with especially unfavorable outcomes when experienced closeness is lower than desired \parencite{pusch2022closeness}. Population-based evidence further indicates that wanting more closeness is more strongly associated with poorer relationship characteristics than experiencing more closeness than desired \parencite{pietras2024closeness}.

These findings support the construct validity of a perceived--desired discrepancy, but they do not establish the ARC-specific causal-temporal claim. Most discrepancy studies predict relationship quality, well-being, or longer-term change rather than the next regulatory action within the same individual. H1 is therefore not the claim that closeness discrepancies exist or matter. It is the narrower claim that momentary discrepancy contains incremental information about subsequent regulation after current state, context, and behavioral inertia have been accounted for.

\subsection{Why the difference score must itself compete}

A simple score
\[
e_t=r^\ast_t-\hat r_t
\]
imposes a substantive constraint: desired state and perceived state are assumed to enter the action model with equal magnitude and opposite sign. That constraint should be tested rather than hidden inside data preprocessing.

The H1 study therefore compares a sequence of increasingly informative observation models.

\begin{table}[htbp]
\caption{Competing Observation Models for the H1 Discrepancy Claim}
\label{tab:h1-competing-models}
\centering
\small
\setlength{\tabcolsep}{3pt}
\renewcommand{\arraystretch}{0.95}
\begin{threeparttable}
\begin{tabular}{p{0.09\linewidth}p{0.30\linewidth}p{0.24\linewidth}p{0.24\linewidth}}
\toprule
Model & Additional relational predictors & Question answered & Interpretation \\
\midrule
M0 & None beyond context and prior action & Is generic context plus behavioral inertia sufficient? & Baseline model \\
M1 & Perceived state \(\hat r_t\) & Does current relational perception alone predict action? & Descriptive baseline for relational-state information \\
M2 & \(\hat r_t\) and \(r^\ast_t\) separately & Does desired state contribute beyond perceived state? & Flexible component model; \(\beta_{r^\ast}\) is H1-P1 \\
M3 & Discrepancy \(e_t=r^\ast_t-\hat r_t\) & Does signed discrepancy predict subsequent action? & \(\beta_e\) is H1-P2; equality is checked separately in M2 \\
M4 & Positive and negative discrepancy components & Are closeness frustration and closeness surfeit dynamically asymmetric? & Piecewise discrepancy extension \\
\bottomrule
\end{tabular}
\begin{tablenotes}[flushleft]
\footnotesize
\item Note. All models retain the same preregistered context, elapsed-time, and autoregressive terms. H1-P1 and H1-P2 are direct coefficient tests with Holm family-wise control. Structural contrasts are evaluated separately. Days 12--14 remain untouched for secondary predictive RMSE validation.
\end{tablenotes}
\end{threeparttable}
\end{table}

The discrepancy representation is strengthened if the constrained \(e_t\) model predicts future action at least as well as the less parsimonious unconstrained model containing \(r^\ast_t\) and \(\hat r_t\) separately. It is weakened if perceived state alone performs equally well, because the target adds no useful information, or if the two raw components have materially different effects that are obscured by subtraction.

Existing evidence also makes directional asymmetry plausible. Wanting more closeness than currently experienced may not be dynamically equivalent to wanting less closeness than currently experienced \parencite{pusch2022closeness,pietras2024closeness}. ARC therefore treats positive and negative discrepancy components as a mandatory sensitivity analysis rather than assuming one globally linear slope.

\subsection{Perceived partner responsiveness supports the feedback pathway}

The feedback component also has strong adjacent evidence. In a two-week daily diary study, within-person increases in perceived partner responsiveness covaried with greater emotional expression to the partner; an experimental study in the same program showed that induced responsiveness increased expression of anxiety \parencite{ruan2020responsiveness}. A yearlong intensive longitudinal study further showed that perceived partner responsiveness participated in within- and between-person commitment dynamics and was associated with attachment-related individual differences \parencite{segal2016responsiveness}.

These studies make the ARC feedback pathway plausible, but they also highlight a measurement distinction that the framework must preserve. Perceived responsiveness is not identical to partner behavior. It may represent a psychologically important state estimate while also containing projection, attachment-related interpretation, and common-method variance. Stronger dyadic tests should therefore distinguish actor perception from partner report or independently coded partner behavior whenever possible.

\subsection{Attachment orientations plausibly shape both response magnitude and response channel}

The idea that \Anx\ and \Avo\ modify relational dynamics has direct within-person precedent and is consistent with interactionist accounts in which attachment effects depend on relational context and activation conditions \parencite{campbell2011interactionist}. Event-contingent work found that attachment anxiety strengthened, and attachment avoidance weakened, affective reactivity to agreeable interpersonal behavior, particularly with romantic partners \parencite{sadikaj2011dynamics}. Daily and intensive longitudinal studies likewise show attachment-related differences in relationship feedback, commitment processes, and responses to relational cues \parencite{segal2016responsiveness,beck2024daily,carnelley2025phubbing}. More recent dyadic experience-sampling work using dynamic structural equation modeling reports attachment-related moderation of emotional interdependence between partners \parencite{xie2025interdependence}.

This evidence argues against representing attachment only as a shift in one signed action coefficient. Anxious regulation can be expressed through overt proximity seeking, reassurance seeking, protest, or heightened interpersonal engagement, while avoidant regulation can appear as weaker overt responding, distancing, suppression, or internal disengagement. The same relational discrepancy can therefore be translated into different response channels.

ARC consequently treats H2 as a multivariate policy claim. The key test is whether \Anx\ and \Avo\ explain reproducible heterogeneity in channel-specific discrepancy-to-action mappings after context and action history are controlled. The strongest directional prediction concerns positive discrepancy, where a person wants more closeness than is currently experienced: higher \Anx\ should preferentially strengthen overt approach-oriented responses, whereas higher \Avo\ should reduce overt approach and increase distancing or internal disengagement. Effects for closeness surfeit remain less constrained and should be estimated rather than assumed.

This literature makes H2 plausible, but does not establish the ARC-specific \(e_t\rightarrow a_{t+1}\) moderation. H2 therefore remains contingent on H1 measurement first proving viable.

\subsection{Secure-base processes justify a candidate, not yet a required core construct}

Adult secure-base functioning can be measured behaviorally. In engaged couples, the Secure Base Scoring System captured secure-base use and support in observed problem-solving interactions and predicted relationship variables beyond a generic communication coding system \parencite{crowell2002securebase}. Daily dyadic work likewise found reciprocal associations between relationship-specific security and support seeking, provision, receipt, and felt support \parencite{davila2009securebase}.

There is also direct evidence that attachment is associated with recovery from relational stress. Cortisol trajectories after relationship conflict vary with attachment insecurity \parencite{powers2006cortisol,pietromonaco2013cortisol}, positive dyadic coping predicts faster cortisol recovery and interacts with attachment anxiety \parencite{meuwly2012cortisol}, and observed recovery from romantic conflict has developmental links to earlier attachment security and partner functioning \parencite{salvatore2011recovery}. Coregulatory and partner-buffering accounts independently place interpersonal regulation and recovery at the center of adult attachment processes \parencite{sbarra2008coregulation,simpson2014buffering,overall2022buffering}. Relationship experiences can also contribute to longer-term changes in attachment security, supporting a distinction between momentary regulation and slower representational change \parencite{arriaga2019security,bayraktaroglu2023positive}.

However, this evidence does not yet demonstrate the specific ARC H3 claim. Much of the recovery literature operationalizes security through low insecurity, attachment classifications, developmental history, or partner behavior. The direct incremental question remains open: does a positive secure-base representation \Sec\ predict recovery after \Anx\ and \Avo\ are already modeled?

H3 is therefore deliberately stronger than the existing evidence and serves as a promotion test rather than as an assumption required to define ARC. \Sec\ should be promoted to the person-level core only if a positive measure of secure-base representation improves a preregistered primary recovery outcome beyond low attachment anxiety and avoidance, adds genuinely out-of-sample predictive information, and the increment is independently replicated or supported by a second preregistered design. The hierarchy of tests is summarized in Table~\ref{tab:h2-h3-tests}.

\begin{table}[htbp]
\caption{Predefined Tests for Attachment-Dependent Policy and Secure-Base Recovery}
\label{tab:h2-h3-tests}
\centering
\scriptsize
\setlength{\tabcolsep}{2pt}
\renewcommand{\arraystretch}{0.85}
\begin{threeparttable}
\begin{tabular}{p{0.09\linewidth}p{0.26\linewidth}p{0.27\linewidth}p{0.26\linewidth}}
\toprule
Test & Model contrast & Target quantity & Decision rule \\
\midrule
H2a & Channel-specific action model without versus with \(\Anx \times e_t\) & Approach/proximity-seeking response to closeness deficit & Retain only if moderation is reproducible and improves held-out prediction \\
H2b & Channel-specific action model without versus with \(\Avo \times e_t\) & Overt approach, distancing, and internal disengagement & Retain channel-specific effects; do not force one signed-action interpretation \\
H2c & Actor-only versus actor--partner moderation & Partner attachment contribution to actor response & Treat as dyadic extension unless incremental fit and prediction justify it \\
H3a & Recovery baseline versus baseline plus \(\Anx,\Avo\) & Recovery explained by insecurity dimensions & Establishes comparison model, not positive security \\
H3b & H3a plus \(\Sec\) & Incremental value of positive secure-base representation & Promotion requires primary-outcome improvement beyond \(\Anx,\Avo\), out-of-sample value, and replication; target retreat alone is not recovery \\
H3c & H3b plus observed/perceived partner support & Whether \(\Sec\) remains informative after proximal support is modeled & Stronger evidence for independent core status; otherwise report conditional attenuation, not identified mediation \\
\bottomrule
\end{tabular}
\begin{tablenotes}[flushleft]
\footnotesize
\item Note. H2 is evaluated on separate action channels as well as any signed summary. H3 is a promotion test for a positive secure-base measure; secondary recovery outcomes cannot rescue failure on the preregistered primary endpoint.
\end{tablenotes}
\end{threeparttable}
\end{table}

\subsection{Core hypotheses and candidate-core promotion test}

\textbf{H1: Discrepancy--action coupling.} Within individuals, measured desired-minus-perceived relational discrepancy at time \(t\) predicts subsequent regulatory action at \(t+1\) beyond measured context, perceived state alone, and autoregressive action history.

\textbf{H2: Attachment-dependent policy.} Attachment anxiety and avoidance explain reproducible between-person differences in how discrepancy and context are translated into specific regulatory channels. Anxiety is expected to preferentially strengthen overt approach-oriented responses to closeness deficit, whereas avoidance is expected to reduce overt approach and increase distancing or internal disengagement. The prediction concerns moderation of measured temporal relations, not assignment of fixed controller types.

\textbf{H3: Secure-base promotion test.} Positive secure-base representation predicts more effective recovery and return to constructive activity beyond what is explained by low anxiety and avoidance and, in stronger tests, beyond proximal partner support. H3 is the preregistered criterion for promoting \Sec\ from candidate-core status to the required person-level core.

\subsection{Extension hypotheses}

\textbf{H4: Candidate moderator specificity.} Mentalized affectivity predicts or moderates externally scored relational-state inference quality, while emotion-regulation flexibility predicts or moderates context- and feedback-sensitive action adaptation. These effects must be incremental to adjacent constructs and are not required for the core ARC loop.

\textbf{H5: Dynamic organization and external disorganization.} Preregistered dynamic indicators such as poor recovery, context-residual inconsistency, persistent approach--avoidance conflict, or maladaptive switching improve held-out prediction of independently measured adult disorganization beyond \Anx, \Avo, their interaction, and simpler static summaries.

\subsection{Evidence hierarchy}

The five hypotheses do not have equal evidential status. H1 is the necessary first test because the feedback architecture loses its empirical purpose if measured discrepancy does not predict subsequent action. H2 tests whether established attachment dimensions modulate the policy. H3 is a candidate-core promotion test for positive security, not a requirement for the current minimal model. H4 and H5 are extensions whose failure should simplify the framework rather than trigger redefinition of the core.

\begin{table}[htbp]
\caption{Evidence Status and Decisive Tests for ARC Hypotheses}
\label{tab:hypothesis-evidence}
\centering
\scriptsize
\setlength{\tabcolsep}{2pt}
\renewcommand{\arraystretch}{0.85}
\begin{threeparttable}
\begin{tabular}{p{0.08\linewidth}p{0.27\linewidth}p{0.27\linewidth}p{0.26\linewidth}}
\toprule
Hypothesis & Adjacent empirical support & ARC-specific evidence still missing & Decisive result \\
\midrule
H1 & Closeness discrepancies predict relational outcomes \parencite{frost2013closeness,pusch2022closeness,pietras2024closeness} & Within-person lag from measured discrepancy to subsequent regulatory action & Target/state information improves held-out action prediction and the retained discrepancy representation survives comparison with simpler alternatives \\
H2 & Attachment moderates within-person reactivity, daily relational responses, and dyadic interdependence \parencite{sadikaj2011dynamics,segal2016responsiveness,xie2025interdependence,carnelley2025phubbing} & Direct moderation of channel-specific \(e_t\rightarrow a^{(k)}_{t+1}\) mappings & Reproducible cross-level moderation of approach, distance, or internal regulation after H1 measurement is established \\
H3 candidate core & Adult secure-base behavior is measurable; attachment and dyadic coping predict support and stress recovery \parencite{crowell2002securebase,davila2009securebase,powers2006cortisol,meuwly2012cortisol,salvatore2011recovery} & Incremental prediction of preregistered recovery outcomes by positive \(S\) beyond \(L/U\) and, in stronger tests, proximal support & Promote \(S\) only after preregistered primary-outcome improvement beyond insecurity, out-of-sample value, and replication \\
H4 & Mentalized affectivity and flexibility have construct-level support \parencite{greenberg2017mentalized,mosannenzadeh2024flexibility} & Pathway-specific incremental validity in repeated relational data & Moderator improves its assigned pathway beyond adjacent constructs \\
H5 & Adult disorganization is measurable but heterogeneous \parencite{pollard2023disorganized} & Incremental validity of dynamic trajectory markers & Dynamic markers predict an independent criterion beyond static attachment baselines \\
\bottomrule
\end{tabular}
\begin{tablenotes}[flushleft]
\footnotesize
\item Note. Adjacent support is not treated as confirmation of the ARC-specific temporal hypothesis.
\end{tablenotes}
\end{threeparttable}
\end{table}

\subsection{Comparison with prior models}

\begin{table}[htbp]
\caption{Selected Precedents and the Empirical Position of ARC}
\label{tab:model-comparison}
\centering
\scriptsize
\setlength{\tabcolsep}{2pt}
\renewcommand{\arraystretch}{0.85}
\begin{threeparttable}
\begin{tabular}{p{0.17\linewidth}p{0.23\linewidth}p{0.23\linewidth}p{0.25\linewidth}}
\toprule
Approach & Existing contribution & Human dynamic evidence & ARC empirical role \\
\midrule
Bowlby / Kobak & Goal-directed attachment control and emotion-regulation set-goals & Observational mother--teen work for Kobak et al. & Does not claim novelty; translates control assumptions into repeated adult measures \\
Stevens and Zhang & Continuous dynamical attachment system with recovery-like behavior & Simulation / parameter analysis & Psychological rather than neurochemical state measurement in adults \\
Petters & Explicit computational lineage of the attachment control system & Primarily theoretical / computational & Uses the lineage as prior theory, not as an ARC-specific mechanism \\
Gagliardi & Multidimensional set-goals, perception, action, feedback, and agent interaction & Computational implementation & Tests a comparable temporal logic with intensive human observations \\
Murray / Sbarra and Hazan & Risk regulation, coregulation, separation and recovery in adult relationships & Human relationship literature and empirical agenda & Integrates adult interpersonal regulation with explicit time-ordered measurement \\
Olar et al. & Adaptive network account of anxious--avoidant dyadic dynamics & Computational model & Provides a rival dynamic architecture for comparison \\
ARC & Parsimonious sequence from measured state and target to action, feedback, and recovery & Proposed intensive longitudinal and dyadic tests & Contribution is integration, measurement architecture, held-out prediction, and falsification \\
\bottomrule
\end{tabular}
\begin{tablenotes}[flushleft]
\footnotesize
\item Note. The table identifies differences in scope and evidential role. It is not a ranking of models.
\end{tablenotes}
\end{threeparttable}
\end{table}

The comparison emphasizes that ARC does not originate control-system attachment. Its proposed addition is a common measurement bridge from person-level attachment variables to repeated relational states, actions, partner feedback, and recovery in adult relationships.

\section{Measurement and Methods}

\subsection{Stage 1: confirmatory feasibility study for H1}

The first human study should test the minimum temporal claim before expanding ARC. The temporal design comprises 14 consecutive days with six synchronized, semi-random prompts per day, yielding 84 planned observations per participant. The earlier 120{,}000-replication exact-design Monte Carlo selected 84 cohabiting romantic couples as the smallest tested candidate robust to the then-current confirmatory lag-sufficiency criterion at approximately 75\% prompt compliance under the prespecified informative-missingness stress condition. Because the confirmatory analysis now excludes the non-comparable prompt-1-to-prompt-2 transition, the final analysis target is being recalibrated prospectively across 72, 84, 96, 108, and 120 couples before recruitment is frozen. The recruitment ceiling remains a separate pre-recruitment operational decision because couple-level attrition was not calibrated by either simulation. Enrollment may not be extended in response to observed H1 coefficients, structural classifications, or holdout performance.

The proposed design is intentionally denser than a once-daily diary. Recent simulation work on dynamic structural equation models indicates that satisfactory estimation depends jointly on the number of individuals and the number of repeated observations; configurations around at least 100 individuals and 30 or more time points have performed adequately for common DSEM targets, while more complex within-level moderation requires larger \(N\), larger \(T\), or both \parencite{mcneish2020dsem,fang2024missing,wang2026dsem,holtmann2025latent}. The ARC-specific exact-design simulation provides the direct planning basis for this study: the design assumes prompt compliance of approximately 75\% or higher, whereas none of the tested candidate sample sizes was robust to the same lag-sufficiency rule at sustained 65\% compliance.

\subsection{Temporal ordering and the H1 estimands}

At prompt \(t\), participants first report perceived relational state \(\hat{\mathbf r}_{i,t}\) and desired relational state \(\mathbf r^\ast_{i,t}\) on matched coordinates. The primary pilot coordinate is closeness because the desired and experienced state can be expressed on the same response scale. The discrepancy is calculated as
\[
\mathbf e_{i,t}=\mathbf r^\ast_{i,t}-\hat{\mathbf r}_{i,t}.
\]

At prompt \(t+1\), participants report regulatory actions that occurred since the previous prompt. This preserves the intended temporal order from discrepancy measured at \(t\) to action occurring during the subsequent interval. Overnight transitions and the within-day prompt-1-to-prompt-2 transition are excluded from the primary lagged analysis because the source action at the first prompt does not summarize a standardized preceding same-day ESM interval.

The primary action outcome is a signed relational-action score,
\[
a^{\mathrm{signed}}_{i,t+1}
=
a^{\mathrm{approach}}_{i,t+1}
-
a^{\mathrm{distance}}_{i,t+1},
\]
while the approach and distance components are retained separately for mandatory sensitivity analyses. Internal regulation, such as suppression or cognitive distancing without overt relational action, is measured separately rather than forced into the signed interpersonal-action score.

The primary H1 test is coefficient-first. H1-P1 evaluates the within-person desired-state coefficient in the unconstrained M2 model after adjustment for perceived state, context, and action history. H1-P2 evaluates the within-person signed-discrepancy coefficient in M3. The two tests form one Holm-controlled confirmatory family and directly test whether each coefficient exceeds a preregistered smallest effect size of interest, rather than first testing against zero and then applying a separate relevance filter. Structural representation is characterized separately through the M2 symmetry contrast and the M4 positive--negative discrepancy contrast.

All time-varying relational predictors are decomposed into within-person deviations and participant means. Participant-paired RMSE on the locked days 12--14 temporal holdout is retained as secondary predictive evidence rather than a sequential model-selection gate. For example, the signed-discrepancy model is
\begin{equation}
a^{\mathrm{signed}}_{i,t+1}
=
\alpha
+
\beta_e e^{\mathrm{WP}}_{i,t}
+
\beta_{\bar e}\bar e_i
+
\phi a^{\mathrm{signed,WP}}_{i,t}
+
\phi_B\bar a_i
+
\boldsymbol{\beta}_c^\top \mathbf c_{i,t}
+
\epsilon_{i,t+1},
\label{eq:h1-pilot}
\end{equation}
where \(\mathbf c_{i,t}\) contains the locked context terms and elapsed time. A positive \(\beta_e\) is one of the two co-primary H1 claims. Its interpretation depends on multiplicity-controlled uncertainty and the preregistered effect-size threshold, not on a preceding M1-versus-M0 prediction gate.

\subsection{The discrepancy score is a testable restriction}

The primary discrepancy model is not compared only against a null model. Because \(e_t=r^\ast_t-\hat r_t\) constrains target and perception to have equal-and-opposite effects, the analysis explicitly compares the model hierarchy in Table~\ref{tab:h1-competing-models}. In particular, M2 estimates perceived and desired closeness separately:
\begin{equation}
a^{\mathrm{signed}}_{i,t+1}
=
\alpha_i
+
\beta_{\hat r}\hat r^{\mathrm{WP}}_{i,t}
+
\beta_{r^\ast}r^{\ast,\mathrm{WP}}_{i,t}
+
\phi_i a^{\mathrm{signed}}_{i,t}
+
\boldsymbol{\beta}_c^\top\mathbf c_{i,t}
+
\epsilon_{i,t+1}.
\label{eq:h1-components}
\end{equation}

The simple discrepancy representation corresponds to the restriction \(\beta_{r^\ast}=-\beta_{\hat r}\), equivalently \(d_{\mathrm{sym}}=\beta_{r^\ast}+\beta_{\hat r}=0\). The restriction is evaluated directly with a preregistered equivalence tolerance. Held-out M3-versus-M2 prediction is reported as convergent or discordant secondary evidence, but predictive non-inferiority is not used as a gate for whether the structural restriction may be tested.

A second sensitivity model decomposes
\[
e^+_t=\max(e_t,0),
\qquad
e^-_t=\max(-e_t,0),
\]
because closeness frustration and closeness surfeit need not regulate symmetrically \parencite{pusch2022closeness,pietras2024closeness}. This piecewise model is an extension of H1 rather than a replacement chosen after inspecting the data.

\subsection{Prompt schedule}

Each couple receives six synchronized prompts per day across a preregistered waking interval. The schedule is semi-randomized within daily strata and enforces minimum spacing so that repeated observations remain intensive without becoming effectively continuous. Prompts expire after a preregistered response window, and confirmatory lags use actual response timestamps rather than nominal prompt order. Exact scheduling parameters, reminder timing, validity rules, and implementation details are maintained in the study protocol and reproducibility package rather than treated as theoretical claims in the main text.

This design is intended to produce enough same-day lag pairs for a dynamic analysis while avoiding excessive burden. The pilot protocol therefore excludes physiological sensing, lengthy momentary questionnaires, and disorganization measures from the primary H1 test.

\subsection{Person-level measures}

Attachment anxiety and avoidance are measured at baseline with a validated dimensional instrument such as the ECR-R \parencite{fraley2000irt,sibley2005ecrr}. These measures describe the sample and support exploratory H2 analyses, but they are not needed to establish H1.

Positive secure-base representation is not required for the H1 confirmatory analysis. It may be collected for feasibility work toward H3 only if a validated implementation and scoring workflow are available \parencite{waters2006workingmodels,waters2019securebase}. Mentalized affectivity and emotion-regulation flexibility are optional and should not be added merely to maximize the number of ARC constructs measured \parencite{greenberg2017mentalized,mosannenzadeh2024flexibility}.

\subsection{Momentary measurement}

The H1 pilot prioritizes a small set of repeatedly measurable variables:

\begin{enumerate}[label=(\arabic*)]
\item current perceived closeness to the partner, \(\hat r_t\);
\item current desired closeness, \(r^\ast_t\), on the identical response scale;
\item approach or proximity-seeking action since the previous prompt;
\item overt distancing or space-seeking action since the previous prompt, without treating it as a direct measure of autonomy;
\item internal suppression or disengagement since the previous prompt;
\item perceived partner responsiveness or availability since the previous prompt;
\item negative affect or attachment-relevant distress;
\item conflict or salient relational stressor since the previous prompt;
\item interaction/contact context and elapsed time.
\end{enumerate}

The exact item wording is treated as a measurement instrument and must be cognitively piloted before confirmatory data collection. If a Polish-language study is conducted, translation and adaptation should be completed before preregistration rather than after inspection of the outcome data.

Repeatedly asking about both perceived and desired relational states may itself alter attention to relational discrepancies or induce target drift. Measurement reactivity is therefore treated as a design threat rather than assumed away. The pilot should preregister a descriptive reactivity diagnostic that tests whether perceived state, desired state, or signed discrepancy shows a systematic study-day trend after accounting for time-of-day structure and other locked covariates. A monotonic trend is not by itself proof of reactivity, because genuine relationship change can produce the same pattern, but a pronounced trend would motivate sensitivity analyses with study day and day-by-discrepancy terms and would constrain interpretation of later-wave dynamics.

\subsection{Validation architecture}

H1 has both inferential and predictive components, but the confirmatory claim is primarily inferential. The locked M0--M4 family is fit using days 1--11. H1-P1 and H1-P2 are evaluated directly from the preregistered within-person coefficients with Holm multiplicity control and an effect-size threshold. M2/M3/M4 structural quantities are reported independently rather than collapsed into a mandatory winning representation. The M2 symmetry restriction is evaluated against the preregistered equivalence region. A focused N=84 Monte Carlo showed that the M4 asymmetry contrast is too imprecise for a defensible confirmatory practical-equivalence threshold in this pilot, so \(d_{\mathrm{asym}}\) is retained as secondary descriptive structural evidence with its confidence interval. Participant-paired RMSE on days 12--14 provides an untouched secondary test of temporal generalization.

Person-level means used in within/between decomposition are estimated from observed training-period prompts only. Holdout observations therefore cannot alter centering, feature construction, or between-person summaries. This prevents a subtle form of temporal leakage that would otherwise allow days 12--14 to influence their own predictors.

Model specification, transformations, lag definitions, context terms, estimator settings, and the predictive comparison margin are fixed before the test period is opened. The current pilot evaluates later observations within known participants; generalization to entirely held-out participants is a distinct, stronger validation target for a later study.

Missingness, compliance, and the number of valid same-day lag pairs are treated as design outcomes rather than nuisance details. A study that fails to retain enough repeated observations for the planned model is classified as a feasibility failure rather than interpreted as evidence against H1.

\subsection{Missing-data sensitivity}

The primary analysis uses observed valid same-day lags and never reconstructs missed prompts. Because ordinary GEE can be sensitive to informative observation processes, a prespecified sensitivity analysis uses stabilized inverse-probability-of-observation weights estimated from response history \parencite{chen2010wgee,cole2008ipw}. Both the numerator and denominator response models are fit using days 1--11 only, and the training-derived model and truncation bounds are then applied unchanged to days 12--14.

The denominator model uses only information available before each scheduled prompt, including prompt stratum, study day, prior response rate, immediately previous response status, and last observed negative affect, absolute discrepancy, and absolute signed action. The current unobserved prompt state is never used to predict its own availability. Weight distributions, fitted response probabilities, truncation frequency, and effective sample size are reported.

This weighted analysis is interpreted as a missing-at-random sensitivity conditional on the specified observed history. It is not claimed to identify or correct a missing-not-at-random mechanism in which response depends on the contemporaneous unobserved relational state or action. The exact-design Monte Carlo deliberately retains such a contemporaneous state/action-dependent nonresponse condition as a stress test. Robustness across unweighted and weighted analyses is informative, but agreement between them does not establish that missingness is MAR.

\subsection{Analytic progression}

The confirmatory H1 estimator is a Gaussian generalized estimating equation with identity link, couple-level clustering, an independence working correlation, and a bias-reduced sandwich covariance estimator \parencite{zeger1986gee,mancl2001gee}. The mean model contains explicit within-person deviations and participant means, the locked context terms, prompt stratum, elapsed time, and the relevant M0--M4 relational predictors. This estimates population-average temporal associations while allowing repeated observations from both partners in a couple to remain statistically dependent.

The choice is deliberately narrower than a full latent dynamic model. H1 asks whether the average state/target architecture adds predictive information and which observation representation is warranted; it does not require estimation of a person-specific latent controller. Couple and person random intercepts and relational-slope heterogeneity are nevertheless included in the Monte Carlo data generator. The confirmatory GEE must therefore demonstrate acceptable bias, interval coverage, false-positive behavior, and predictive model selection under richer heterogeneity than it explicitly parameterizes.

The working correlation is set to independence so that the within-couple covariance structure is not selected from the data. Robust sandwich inference is used for correlation misspecification, with a finite-cluster correction prespecified for primary inference \parencite{zeger1986gee,mancl2001gee}. Software versions, convergence tolerances, iteration limits, and Monte Carlo execution parameters are frozen in the reproducibility package and study protocol rather than repeated in the theoretical manuscript. There are no Bayesian priors and no required random-effects distribution in the primary H1 estimator.

DSEM, multilevel mixed models, actor--partner interdependence models, continuous-time state-space models, and multilevel vector autoregression remain scientifically useful sensitivity or later-stage models \parencite{mcneish2020dsem,wang2026dsem,holtmann2025latent}. In particular, H2 requires explicit study of random-slope and cross-level moderation and should not inherit the simpler H1 estimator by default.

\subsection{Stage 2: attachment-dependent policy}

H2 should not be tested by adding one omnibus attachment interaction to the signed H1 outcome. The action channels are modeled separately:
\[
a^{\mathrm{approach}}_{t+1},\quad
a^{\mathrm{distance}}_{t+1},\quad
a^{\mathrm{internal}}_{t+1}.
\]
The confirmatory H2 model adds cross-level interactions between person-level \Anx/\Avo\ and within-person discrepancy to otherwise identical action equations. This requires more between-person information than H1, so the H1 pilot should not be retroactively treated as a definitive H2 test merely because ECR-R scores are available.

The Stage 2 design should therefore be powered specifically for random-slope heterogeneity and cross-level moderation. At minimum it should preregister:
\begin{enumerate}[label=(\arabic*)]
\item which discrepancy representation survived H1 model comparison;
\item the action channel or channels that are confirmatory;
\item directional predictions for \Anx\ and \Avo\ by discrepancy sign;
\item whether actor and partner attachment effects are both modeled;
\item how multiple channel-specific interactions are controlled or hierarchically modeled;
\item a design-specific simulation for the interaction effect.
\end{enumerate}

A dyadic extension can include partner attachment and partner behavior, but actor-only moderation should remain distinguishable from actor--partner interaction effects. Existing dynamic dyadic studies demonstrate that such moderation is estimable, while also showing that sex/gender, partner role, and modest sample size can complicate interpretation \parencite{xie2025interdependence}.

\subsection{Stage 2b: candidate-core promotion through secure-base recovery}

H3 requires a design that actually contains recoverable perturbations. An ESM study without identifiable stress episodes can estimate ordinary discrepancy dynamics but may provide weak information about recovery time. Two complementary designs are therefore appropriate.

\textbf{Ecological recovery.} Event-contingent sampling identifies relationship conflict, perceived rejection, separation, or another preregistered attachment-relevant perturbation. Recovery is then quantified from subsequent prompts using discrepancy burden, time to functional tolerance, return to constructive activity, and target-versus-state change.

\textbf{Laboratory recovery.} A standardized dyadic stress or conflict paradigm is followed by a prespecified recovery period. Behavioral recovery can be coded directly, and physiological outcomes such as cortisol can be added when resources permit. Prior studies demonstrate that attachment and dyadic coping are associated with observed and physiological recovery after interpersonal stress \parencite{powers2006cortisol,meuwly2012cortisol,salvatore2011recovery}.

The H3 candidate \Sec\ must be a positive secure-base representation or secure-base construct measured independently of the outcome. Secure-base behavior observed in the recovery episode is better treated as a proximal process or validation variable than as the same person-level \Sec\ predictor \parencite{crowell2002securebase}. The decisive promotion comparison is whether adding \Sec\ improves a preregistered primary recovery outcome after \Anx\ and \Avo\ are already in the model and improves genuinely out-of-sample prediction. Failure of this comparison leaves \Sec\ outside the required ARC core. A study must nominate one primary recovery endpoint before data inspection. The endpoint must not treat a reduction in \(|e_t|\) caused solely by retreat of \(r_t^\ast\) as successful recovery. In ecological ESM, discrepancy burden may be used only together with the preregistered decomposition of target change and perceived-state change; an outcome such as return to constructive activity provides a useful criterion that is not algebraically determined by the discrepancy score. Laboratory studies may instead nominate a behavioral or physiological recovery endpoint. Secondary outcomes test convergence rather than provide alternative opportunities for promotion.

A stronger model subsequently adds observed or perceived partner support. This tests whether \Sec\ captures person-level representational information beyond the support available in the episode. If only proximal support predicts recovery, ARC should not retain a separate \Sec\ effect without evidence.

\subsection{Stage 3: dynamic organization and external disorganization}

Stage 3 evaluates candidate dynamic indicators of disorganization against independent criteria only after the core temporal and recovery architecture is viable. Interview, behavioral, and self-report disorganization measures are treated as different external criteria rather than assumed to be interchangeable \parencite{pollard2023disorganized}.

The full H1 pilot protocol, preregistration draft, data dictionary, and design-simulation specification are maintained under \texttt{studies/pilot\_h1/}.

\section{Falsification and Simplification Criteria}

ARC is intended to become smaller when evidence does not support added structure.

\subsection{Core temporal sequence}

The current H1 pilot separates two coefficient-level claims from structural and predictive evidence. H1-P1 tests the desired-state coefficient in M2 after adjustment for perceived state, context, and action history. H1-P2 tests the signed-discrepancy coefficient in M3. Both use the locked one-sided SESOI tests with Holm family-wise control. H1-P2 alone does not establish a contribution beyond perceived state: M3 does not include a separate perceived-state coefficient, and the difference score contains perceived-state information. Even joint support does not uniquely identify discrepancy regulation as a psychological mechanism.

The target component should be simplified if adequately powered studies repeatedly find negligible within-person target variation and no incremental target information. A nonsignificant coefficient or an imprecise prediction difference in one pilot is insufficient evidence of absence. If target drift is substantial, it must be modeled explicitly rather than treated as error. More consistent reverse or simultaneous associations would challenge the proposed temporal ordering.

The difference-score representation is a testable restriction. For M2, report \(d_{\mathrm{sym}}=\beta_{r^\ast}+\beta_{\hat r}\) with the prespecified 90\% confidence interval and tolerance \(\delta_{\mathrm{sym}}=1.0\) on the coefficient-per-10-point scale. An interval wholly inside \([-1,1]\) supports practical compatibility with the restriction; an interval wholly outside it on one side provides evidence against it; overlap is inconclusive. Compatibility within a tolerance does not establish exact equality or a unique mechanism.

For M4, report the positive and negative discrepancy slopes and \(d_{\mathrm{asym}}=\beta_{+}+\beta_{-}\), where \(\beta_{-}\) multiplies \(\max(-e,0)\). Report the estimate and 95\% confidence interval descriptively. The focused validation did not justify a confirmatory asymmetry tolerance at \(N=84\). Therefore, this pilot does not classify M4 as practically symmetric or asymmetric, and an interval containing zero is not evidence of symmetry.

Participant-paired temporal-holdout RMSE differences and their uncertainty are secondary predictive evidence. They neither gate the co-primary tests nor impose a mandatory winning model. Reproducible evidence against the M2 coefficient restriction motivates separate state and target pathways; replicated asymmetric patterns with adequate precision motivate a prospectively calibrated asymmetric extension. Structural incompatibility with little predictive loss means that the simple representation may remain a useful approximation, not that its psychological mechanism has been established. Discordant or imprecise evidence must remain visible rather than be converted into a favorable model choice.

\subsection{Attachment moderators}

The dynamic role assigned to \Anx\ and \Avo\ should be removed if they do not reproducibly moderate discrepancy-to-action or related temporal parameters. Their established value as descriptive attachment dimensions would remain; only the ARC-specific dynamic interpretation would fail.

H2 is specifically weakened if attachment predicts only between-person mean levels of approach, distancing, or internal regulation but does not alter within-person discrepancy-to-action mappings. If moderation appears only in a signed action summary but not in its component channels, the channel decomposition should take precedence because opposite high-intensity behaviors can cancel in a signed score.

A null interaction in an underpowered pilot is not sufficient for rejection. H2 should be evaluated only in a design simulated for cross-level moderation and random-slope heterogeneity. Conversely, a statistically detectable interaction that does not replicate or improve held-out prediction should not be retained as an ARC policy parameter.

\subsection{Positive security candidate}

\Sec\ should remain outside the required ARC core unless a positive secure-base measure improves a preregistered primary recovery outcome beyond \Anx\ and \Avo, adds genuinely out-of-sample predictive value, and the increment is independently replicated or supported by a second preregistered design. A zero-order association between \Sec\ and recovery is insufficient evidence for H3 or for promotion to core status. A reduction in discrepancy that is explained only by retreat of the desired state is also insufficient evidence of successful recovery.

If \Sec\ adds information beyond \Anx/\Avo\ but loses that information after proximal partner support or observed dyadic coping is included, ARC should report attenuation conditional on support rather than infer mediation or automatically promote \Sec\ as an independent person-level core construct. Attenuation alone does not identify mediation: shared measurement, confounding, and correlated predictors are alternative explanations. A mediation claim requires a separately specified temporal causal model and defensible identification assumptions. If proximal support predicts recovery while \Sec\ does not, the recovery account should be simplified toward episode-level coregulation.

If \Sec\ predicts relationship quality or secure-base behavior but not the preregistered primary recovery outcome, the positive construct may remain theoretically relevant, but it should remain outside the ARC core and its placement in the dynamic recovery pathway should be removed. Secondary outcomes cannot be substituted post hoc for a failed primary promotion endpoint.

\subsection{Mentalized affectivity and flexibility}

\MentAff\ and \Flex\ should be excluded, merged with adjacent constructs, or redefined if they fail discriminant validity or do not add incremental information to the specific pathway assigned to them. ARC does not depend on preserving these moderators.

\subsection{Dynamic disorganization extension}

Dynamic indicators should not be described as attachment disorganization if they fail to improve held-out prediction of an independent disorganization criterion beyond \Anx, \Avo, their interaction, and simpler static summaries. High variability by itself is insufficient evidence of disorganization.

\subsection{Mechanistic non-uniqueness}

Even successful prediction of the ARC temporal sequence would not uniquely identify a latent computational mechanism. If P/PI/PID, state-space, active-inference, reinforcement-learning, or other models remain observationally equivalent, the defensible conclusion is about temporal prediction and feedback organization rather than a unique psychological algorithm. This identifiability problem is addressed separately in Manuscript 2.

\section{Discussion}

\subsection{What ARC contributes}

ARC is best understood as an empirical integration framework. Control-system attachment, secure-base representations, interpersonal risk regulation, dyadic coregulation, and computational attachment models are established precedents \parencite{bowlby1982attachment,kobak1993control,murray2006risk,sbarra2008coregulation,petters2019control,gagliardi2022control,stevens2009dynamic}. The contribution proposed here is to place these ideas into one parsimonious sequence with variables that can be repeatedly measured in adult romantic relationships and with explicit conditions under which components should be rejected.

This positioning deliberately lowers the novelty claim while strengthening falsifiability. The central question is not whether feedback regulation is a new metaphor for attachment. It is whether measured relational discrepancy and feedback provide reproducible information about subsequent action and recovery beyond static attachment dimensions and generic temporal baselines. Throughout this account, control and controller are cybernetic terms for feedback regulation toward a reference condition; they do not denote interpersonal dominance, coercion, or a controlling personality style.

\subsection{Where the existing human evidence stops}

Several pieces of the proposed loop already have human support. Actual--ideal closeness discrepancies are associated with relational outcomes \parencite{frost2013closeness,pusch2022closeness,pietras2024closeness}; perceived partner responsiveness fluctuates within people and predicts interpersonal expression and relationship dynamics \parencite{ruan2020responsiveness,segal2016responsiveness}; partner-buffering research further shows that relationship partners can attenuate insecurity-related difficulties \parencite{simpson2014buffering,overall2022buffering}; and attachment orientations moderate within-person affective or dyadic responses to relational cues \parencite{sadikaj2011dynamics,beck2024daily,xie2025interdependence,carnelley2025phubbing}.

What is not yet established is the specific ARC chain linking a momentary, explicitly measured desired--perceived discrepancy to subsequent regulatory action and then to partner feedback and state updating. This distinction prevents adjacent evidence from being treated as direct confirmation.

It also motivates a stricter treatment of the discrepancy construct. Difference scores are compact, but they encode equality and sign constraints on the effects of their components. ARC therefore treats the difference score as a parsimonious hypothesis that must compete against the raw components and asymmetric discrepancy models. A failure of the simple difference representation would revise the mathematics without necessarily eliminating the broader feedback-regulation account.

\subsection{Attachment dimensions are policy moderators, not controller labels}

ARC does not interpret anxiety and avoidance as fixed controller types. The current formulation assigns them a narrower empirical role: they may explain why the same momentary discrepancy is translated into different regulatory channels. This distinction matters because overt approach, interpersonal distancing, and internal disengagement are not interchangeable manifestations of one scalar action tendency. Existing within-person and dyadic evidence supports attachment-related heterogeneity in responses to relational cues, and interactionist theory emphasizes that these effects depend on context and activation conditions; the specific ARC cross-level interactions nevertheless remain untested \parencite{campbell2011interactionist,sadikaj2011dynamics,xie2025interdependence,carnelley2025phubbing}.

This formulation also provides a clear failure mode. If \Anx\ and \Avo\ predict only stable mean differences in behavior and do not alter within-person discrepancy-to-action mappings, ARC should retain them as descriptive attachment dimensions while dropping their proposed dynamic policy role.

\subsection{Positive security is a stronger claim than low insecurity}

The secure-base literature supports positive representational and behavioral aspects of security \parencite{waters2006workingmodels,waters2019securebase,crowell2002securebase}. Separate literatures also link attachment and dyadic coping to recovery from conflict and physiological stress \parencite{powers2006cortisol,meuwly2012cortisol,salvatore2011recovery}. These literatures justify keeping \Sec\ as a prespecified candidate rather than dropping positive security from the research program. What remains uncertain is the bridge between these lines: whether a positive secure-base representation predicts recovery after anxiety and avoidance are already modeled.

H3 therefore intentionally adopts an incremental-validity criterion. A zero-order association between \Sec\ and recovery is not sufficient. Nor is evidence that secure partners recover better when security is operationalized only as low insecurity. The positive-security construct earns promotion to the ARC core only if a separately measured \Sec\ construct improves a preregistered primary recovery outcome beyond \Anx/\Avo, adds out-of-sample predictive value, and the increment is independently replicated or supported by a second preregistered design. This predictive criterion does not by itself establish a causal effect of \Sec. Stronger mechanistic interpretation would require experimental or otherwise identification-oriented evidence, and the recovery outcome must distinguish genuine state/function improvement from mere retreat of the desired relational target.

\subsection{Adult romantic relationships as the empirical target}

Much computational attachment work has been developed around infant--caregiver systems or simulation. Adult romantic relationships add different targets, reciprocal agency, relationship-specific histories, explicit communication, autonomy goals, and partner-specific regulation. Autonomy need satisfaction can coexist with relatedness in close relationships rather than representing its simple opposite \parencite{kluwer2020autonomy}. ARC therefore treats dyadic reciprocity as a substantive empirical issue rather than an automatic extension from infant models.

\subsection{Why the framework is deliberately small}

ARC retains only \Anx\ and \Avo\ as required person-level variables, together with the minimal momentary loop needed to test feedback regulation. \Sec\ is a prespecified candidate for promotion after H3, not a required construct. \MentAff\ and \Flex\ are optional moderators. Relational-control propensity is evaluated in a separate psychometric program, while the recoverability of P/PI/PID and other candidate mechanisms is a separate identifiability question.

These neighboring research programs can inform later extensions of ARC but are not prerequisites for the framework itself. This separation prevents a failure of one speculative component or one mechanistic model from being hidden by reinterpretation of another.

\subsection{Security and disorganization}

Positive secure-base representation and recovery provide a way to test whether security contributes information that cannot be reduced to low insecurity. This is an integration of established secure-base ideas with repeated temporal measurement, not a claim that positive security is newly discovered or already required by ARC.

The disorganization extension is more tentative. Poor recovery, context-residual inconsistency, persistent approach--avoidance conflict, and maladaptive switching are candidate trajectory properties, not definitions of adult disorganization. Their value depends on incremental prediction of independent criteria.

\subsection{Limitations}

The framework has no human validation data. Momentary targets may be reactive to measurement, relationship goals may change rapidly, partner effects create reciprocal causation, and the sampling interval can change the apparent dynamics. Repeated prompts that ask participants to compare perceived and desired relational states may increase discrepancy salience, shift the desired target, or alter regulatory behavior. The pilot therefore treats study-day trends in \(\hat{\mathbf r}_t\), \(\mathbf r^\ast_t\), and \(\mathbf e_t\) as reactivity diagnostics and examines whether primary conclusions are sensitive to temporal-trend adjustment. Such trends cannot uniquely identify measurement reactivity, because genuine relational change is observationally similar, so they are interpreted as a limitation and sensitivity check rather than a correction. Measurement error in \(\hat{\mathbf r}_t\) and \(\mathbf r^\ast_t\) directly contaminates discrepancy. Person-specific processes may also differ enough that group-level coefficients conceal important heterogeneity.

A further limitation is observational equivalence. Multiple computational mechanisms can produce similar action trajectories. ARC therefore prioritizes predictive and discriminating evidence over mechanistic labels.

\subsection{Immediate empirical priority}

The next scientific step is a feasibility pilot centered on H1. If perceived and desired relational states cannot be measured reliably on matched coordinates, or their discrepancy does not predict subsequent action beyond context and action history, the model should be simplified before any large-scale test of moderators or disorganization.

\section{Conclusion}

The Attachment Regulation and Control (ARC) framework provides a parsimonious, empirically testable bridge between adult attachment dispositions and moment-to-moment relational regulation. It does not claim to invent feedback control in attachment theory. Its proposed contribution is to integrate established ideas into a temporal measurement architecture for adult romantic relationships:
\[
\hat{\mathbf r}_t,\mathbf r^\ast_t
\rightarrow
\mathbf e_t
\rightarrow
\mathbf a_{t+1}
\rightarrow
\mathbf p_{t+1}
\rightarrow
\hat{\mathbf r}_{t+1}
\rightarrow
\mathrm{recovery}.
\]

The framework makes two levels of claim independently falsifiable. First, desired and perceived relational state must contain information about subsequent regulation beyond context and behavioral inertia. Second, the parsimonious difference representation \(e_t=r^\ast_t-\hat r_t\) must compete successfully with perceived-state-only, unconstrained component, and asymmetric discrepancy alternatives. Failure of the second claim revises the mathematical representation without automatically rescuing or rejecting the broader feedback account.

Attachment anxiety and avoidance are then tested as moderators of channel-specific regulation rather than as fixed controller labels. Positive secure-base representation is deliberately kept outside the required core until H3 demonstrates reproducible incremental recovery value beyond low anxiety and avoidance and, in stronger tests, beyond proximal partner support. Mentalized affectivity and emotion-regulation flexibility remain optional moderators. Adult disorganization remains an external criterion.

ARC earns explanatory value only if these dynamic measurements improve reproducible, held-out prediction beyond simpler alternatives. Its first decisive test is therefore empirical rather than computational: whether repeatedly measured relational states and targets predict subsequent regulatory action in human relationships, and whether the proposed discrepancy representation is actually warranted by the data.

\printbibliography
\end{document}